\documentclass[twocolumn,showpacs,preprintnumbers,amsmath,amssymb]{revtex4}
\usepackage{dcolumn}
\usepackage{bm}
\usepackage{amsmath,amssymb,amsfonts,latexsym,fancyhdr,graphicx}

\newcommand{\ket}[1]{\displaystyle{|#1\rangle}}
\newcommand{\bra}[1]{\displaystyle{\langle#1|}}

\newcommand{\Om}{\Omega}
\newcommand{\om}{\omega}

\newcommand{\G}{\Gamma}
\newcommand{\g}{\gamma}
\newcommand{\s}{\sigma}
\newcommand{\la}{\lambda}

\begin{document}
\title{Pseudomodes as an effective description of memory}
\author{L. Mazzola,$^1$ S. Maniscalco,$^1$ J. Piilo,$^1$ K.\,-\,A. Suominen,$^1$ and B. M. Garraway$^2$}
\affiliation{$^1$Department of Physics and Astronomy, University of
Turku, FI-20014 Turun yliopisto, Finland\\$^2$Department of Physics
and Astronomy, University of Sussex, Falmer, Brighton, BN1 9QH,
United Kingdom}

\begin{abstract}
We investigate the non\,-\,Markovian dynamics of two\,-\,state
systems in structured reservoirs. We establish a connection between
two theoretical quantum approaches, the pseudomodes [B. M. Garraway,
Phys. Rev. A \textbf{55}, 2290 (1997)] and the recently developed
non\,-\,Markovian quantum jump method [J. Piilo \textit{et al.},
Phys. Rev. Lett. \textbf{100}, 180402 (2008)]. This connection
provides a clear physical picture of how the structured reservoir
affects the system dynamics, suggesting the role of the pseudomodes
as effective description of environmental memory.
\end{abstract}
\maketitle

\section{Introduction}
Recent advances in experimental techniques for coherent control of
small quantum systems have paved the way to a series of spectacular
experiments aimed at both testing fundamental features of quantum
theory and implementing logic gates for quantum information
processing \cite{Harbook, qinfproc}. Quantum properties, however,
are very fragile. Any interaction between quantum systems and their
surroundings gives rise to decoherence and dissipation phenomena,
destroying the quantumness of the state of the system. For this
reason during the last decade several theoretical and experimental
studies have been devoted to the investigation of the dynamics of
open quantum systems \cite{Breuer, Gardiner, Weiss}.

One of the approaches to the description of open quantum systems
consists in separating the total system into two parts, the quantum
system of interest and the surrounding environment. The environment
is often modeled as an infinite collection of quantum harmonic
oscillators in thermal equilibrium \cite{Breuer}. This type of
reservoir describes, e.g., the quantized electromagnetic field. One
of the key quantities characterizing the reservoir, and therefore
determining the open system dynamics, is the spectral distribution
or structure function. This quantity describes the
frequency\,-\,dependent coupling between the system and the
continuum of harmonic oscillators forming the environment.

In many physical situations the system\,-\,reservoir coupling
strength does not depend strongly on the frequency of the reservoir
oscillators. In this case the reservoir spectral density can be
conveniently approximated by a flat spectrum. One typically refers
to such systems as Markovian open quantum systems. A typical feature
of the dynamics of Markovian open quantum system is the irreversible
flow of energy and/or information from the system to the
environment.

In certain physical contexts, however, the quantum system of
interest interacts with \lq\lq structured\rq\rq reservoirs, whose
spectral density strongly varies with frequency. We refer to such
systems as non\,-\,Markovian open quantum systems. Non\,-\,Markovian
dynamics is characterized by the existence of a memory time scale
during which some energy/information that has been transferred from
the system to the environment feeds back into the system.

Many solid\,-\,state systems such as Josephson junctions, display
strongly non\,-\,Markovian dynamics \cite{Weiss}. Moreover, often
the e.m. field surrounding a quantum system, e.g. an atom, can be
conveniently engineered to prevent or inhibit the occurrence of
decoherent processes such as atomic spontaneous emission. Atoms
placed inside cavities or photonic band gap materials are examples
of quantum systems interacting with such engineered structured
reservoirs \cite{Lambropoulos}. The formalism we develop here for a
two\,-\,level atom can be applied to any two\,-\,level system in
bosonic structured reservoir, e.g, NV vacancy in diamond embedded in
photonic band gap \cite{NVCenters}.

The theoretical description of the dynamics of non\,-\,Markovian
quantum system is usually very complicated, only a few simple
systems are amenable to an exact solution \cite{Breuer}. A number of
methods have been formulated for treating non-Markovian dynamics
\cite{Garrmath, NMQJ, nonMmethods} but the connection between these
methods has so far remained unexplored. Most importantly, due to the
mathematical difficulties in dealing with non\,-\,Markovian systems,
a simple intuitive physical picture of the memory of a
non\,-\,Markovian reservoir and of how such memory allows to partly
restore some of the coherence lost to the environment is highly
desirable. Such a simple effective description is our main result.
By connecting two non\,-\,Markovian approaches, the pseudomodes
\cite{Garrmath} and the non\,-\,Markovian quantum jumps (NMQJ)
methods \cite{NMQJ}, we are able to identify, for simple exemplary
cases, where the memory of a non\,-\,Markovian reservoir resides.

Our results provide insight into the problem of the existence of
pure state trajectories and the physical meaning of the master
equation unravelling for non\,-\,Markovian systems. This issue has
been recently considered in Refs. \cite{Diosi08, Gambetta08}. Both
the pseudomodes and the NMQJ approaches are consistent with the
interpretation given in Ref. \cite{Gambetta08}.

\section{Pseudomode and NMQJ methods}
\subsection{Time-local master equation}
Consider a two\,-\,level atom interacting with a structured
electromagnetic reservoir in the vacuum state. The Hamiltonian of
such a system in the rotating wave approximation is
\begin{equation}\label{Hamiltonian}
H=\hbar\om_{0} \sigma_{+}
\sigma_{-}+\sum_{\la}\hbar\om_{\la}a^{\dagger}_{\la}a_{\la}+\sum_{\la}(\hbar
g^{\ast}_{\la}a_{\la}\s_{+}+\textrm{H.c.})
\end{equation}
where $\s_{\pm}$ are the Pauli raising and lowering operators for
the two\,-\,level system, $\om_{0}$ is the atomic transition
frequency, $a_{\la}$ and $a_{\la}^{\dagger}$ are the annihilation
and creation operators for the mode $\la$ of the field having
frequency $\om_{\la}$ and coupling constant $g_{\la}$. We assume
that initially only one excitation is present in the system. The
state of the total system at time $t$ takes the form
\begin{eqnarray}
\vert \psi(t) \rangle = c_0 \vert g,  0_{\lambda} \rangle + c_1(t)
\vert e,  0_{\lambda} \rangle + \sum_{\lambda} c_{\lambda}(t)\vert g
,1_{\lambda} \rangle,
\end{eqnarray}
with $\vert g, 0_{\lambda} \rangle$, $\vert e,  0_{\lambda}
\rangle$, and $\vert g ,1_{\lambda} \rangle $ the states containing
zero excitations, one atomic excitation and one excitation in the
$\lambda$-mode of the e.m. field, respectively.

The problem of a two\,-\,level system interacting with a zero
temperature reservoir is in principle exactly solvable using Laplace
transforms. The exact non-Markovian master equation describing the
dynamics of the atomic system takes the form \cite{Breuer}
\begin{equation}\label{masteqBre}
\frac{d\rho_{A}}{dt}=\frac{S(t)}{2i}[\s_{+}\s_{-},\rho_{A}]
+\g(t)[\s_{-}\rho_{A}\s_{+}-\frac{1}{2}\{\s_{+}\s_{-},\rho_{A}\}],
\end{equation}
where $\rho_{A}$ is the atomic density operator. The time dependent
Lamb\,-\,shift $S(t)$ and  the time dependent decay rate $\g(t)$ are
given by
\begin{equation}\label{S-g}
S(t)=-2\textrm{Im}\left\{\frac{\dot{c}_{1}(t)}{c_{1}(t)}\right\},\qquad
\g(t)=-2\textrm{Re}\left\{\frac{\dot{c}_{1}(t)}{c_{1}(t)}\right\}.
\end{equation}
The master equation (\ref{masteqBre}) can be simulated by means of
the non\,-\,Markovian quantum jump (NMQJ) method \cite{NMQJ} which
extends the Monte Carlo wave function approach \cite{MCWF} to
non\,-\,Markovian systems with negative decay rates.

\subsection{Pseudomode method}
Alternatively, one can investigate the dynamics using the pseudomode
theory \cite{Garrmath, Garr97}. This method relies on the strong
connection between the atom dynamics and the shape of the reservoir
spectral distribution. More precisely, the key quantities
influencing the time evolution of the atom are the poles of the
spectral distribution in the lower half complex $\om_{\la}$ plane.
By introducing some auxiliary variables, called pseudomodes, defined
in terms of the position and of the residue of the poles of the
spectral distribution, one can derive a Markovian master equation in
the Lindblad form for the extended system comprising the atom and
the pseudomodes. This exact master equation describes the coherent
interaction between the atom and the pseudomodes in presence of
decay of the pseudomodes due to the interaction with a Markovian
reservoir.

\begin{figure}[!]
\begin{center}
\includegraphics[width=8.6cm]{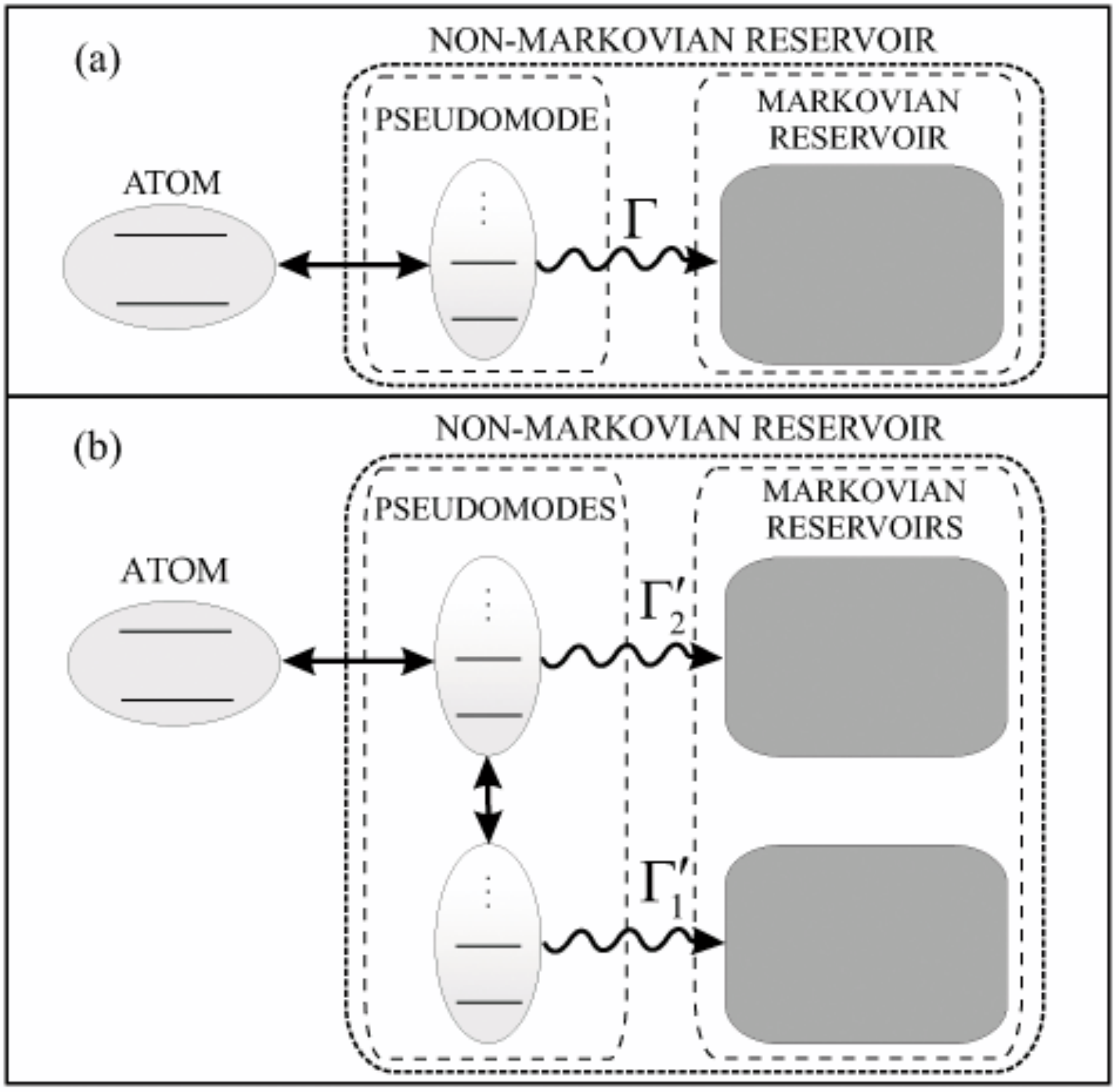}
\end{center}
\caption{Diagramatic representation of the atom\,-\,pseudomode
dynamics. (a) Atom interacting with a Lorentzian structured
reservoir: the atom interacts with a single pseudomode which leaks
into a Markovian reservoir. (b) Atom in a ``simple" photonic band
gap: we see a more complex memory architecture, the second
pseudomode acts as a memory for the atom, while the first pseudomode
acts as a memory for the first one.}\label{fig:schema}
\end{figure}

For a Lorentzian spectral distribution the pseudomode approach leads
to the following master equation
\begin{equation}\label{masteqpseu}
\frac{d\rho}{dt}=-i[H_{0},\rho]-\frac{\G}{2}[a^{\dagger}a\rho-2a\rho
a^{\dagger}+\rho a^{\dagger}a],
\end{equation}
where
\begin{equation}
H_{0}=\om_{0}\s_{+}\s_-+\om_{c}a^{\dagger}a+\Om_{0}[a^{\dagger}\s_{-}+a\s_{+}]
\end{equation}
and $\rho$ is the density operator for the atom and the pseudomode.
Since a Lorentzian function has only one pole in the lower half
complex plane, the atom interacts with one pseudomode only as
displayed in Fig.~\ref{fig:schema}~(a). The constants $\om_{c}$ and
$\G$ are respectively the oscillation frequency and the decay rate
of the pseudomode and they depend on the position of the pole
$z_{1}\equiv\om_{c}-i\G/2$ while $\Om_{0}$ is the pseudomode
coupling constant.

Both master equations \eqref{masteqBre} and \eqref{masteqpseu} are
exact. Hence we expect to obtain an equation of motion for the atom
of the form of Eq.~\eqref{masteqBre} by tracing out the pseudomode
in Eq.~\eqref{masteqpseu}. Additionally it is interesting to see the
expressions of the coefficients $\g(t)$ and $S(t)$ as functions of
the pseudomode amplitude. Indeed the equation obtained from
Eq.~\eqref{masteqpseu} has the form
\begin{equation}\label{tracedME}
\frac{d\rho_{A}}{dt}=\frac{A(t)}{2i}[\s_{+}\s_{-},\rho_{A}]
+B(t)[\s_{-}\rho_{A}\s_{+}-\frac{1}{2}\{\s_{+}\s_{-},\rho_{A}\}],
\end{equation}
where
\begin{equation}\label{Asolution}
A(t)=2 \left[ \om_{0}+\Om_{0}
\frac{\textrm{Re}\{c_{1}(t)b_{1}^{\ast}(t)\}}{|c_{1}(t)|^2} \right],
\end{equation}
and
\begin{equation}\label{Bsolution}
B(t)=2\Om_{0}
\frac{\textrm{Im}\{c_{1}(t)b_{1}^{\ast}(t)\}}{|c_{1}(t)|^2},
\end{equation}
where $b_1(t)$ is the pseudomode amplitude. By using the
differential equations governing the atom\,-\,pseudomode dynamics
\cite{Garrmath}
\begin{equation}\label{diffsyst3}
i\frac{d}{dt}c_{1}=\om_{0}c_{1}+\Om_{0}b_{1},\qquad
i\frac{d}{dt}b_{1}=z_{1}b_{1}+\Om_{0}c_{1},
\end{equation}
it is easy to prove that $A(t)=S(t)$ and $B(t)=\g(t)$.

\subsection{The connection between pseudomode and NMQJ approaches}
Once we have proven the equivalence between the two master
equations, we focus on the simple case of a Lorentzian spectral
distribution off\,-\,resonant with the atomic transition frequency
(damped Jaynes\,-\,Cummings model with detuning).

In the strong coupling regime $\G \ll \Om_0$  the atomic dynamics is
strongly non\,-\,Markovian. The excited state population of the atom
oscillates in time indicating that the energy dissipated into the
environment flows back into the system as a consequence of the
reservoir memory. At the same time the atomic decay rate $\gamma(t)$
attains negative values. The key role of the pseudomode is exposed
when we look at the time derivative of the pseudomode population
and, using the differential equations in Eq.~\eqref{diffsyst3}, we
obtain
\begin{equation}\label{gammapseu1}
\frac{d|b_{1}(t)|^2}{dt}+\G|b_{1}(t)|^2=\g(t)|c_{1}(t)|^2.
\end{equation}
The equation above shows that the \textit{compensated rate of
change} of the pseudomode population, given by the left hand side of
Eq.~\eqref{gammapseu1} (where the effect of the pseudomode leakage
is removed), is directly related to the atomic decay rate. This
means that the information about the dissipative dynamics of the
atom into the structured reservoir are all contained in the
pseudomode dynamics. Equation~\eqref{gammapseu1} and its physical
interpretation is one of the main results of the paper.

We conclude this section analyzing the connection between the NMQJ
unraveling of the master equation in Eq.~\eqref{masteqBre} and the
Monte-Carlo wave function unraveling of the pseudomode master
equation in Eqs.~\eqref{masteqpseu} and \eqref{masteqpseu2}. In the
NMQJ description the ensemble members living in the Hilbert space of
the system are always in a pure state. In particular the density
matrix of the ensemble can be written as
\begin{equation}\label{NMQJunr}
\rho_{A}(t)=\frac{N_{0}(t)}{N}\ket{\psi_{0}(t)}\bra{\psi_{0}(t)}+\frac{N_{1}(t)}{N}\ket{\psi_{1}}\bra{\psi_{1}},
\end{equation}
where
\begin{eqnarray}
&\ket{\psi_{0}(t)}=C_{g}(t)\ket{g}+C_{e}(t)\ket{e},\\
&\ket{\psi_{1}}=\ket{g},
\end{eqnarray}
$N$ is the total number of ensemble members, $N_{1}(t)$ is the
number of ensemble members who have jumped into the ground state,
and $N_{0}(t)$ is the number of members who have not jumped, or
which have gone through a jump-reverse-jump cycle due to the
negative decay rate \cite{NMQJ}.

In the pseudomode description the unraveling of the master equation
\eqref{masteqpseu} is in the extended Hilbert space containing the
pseudomode. The density matrix of such ensemble, expressed in the
atom-pseudomode basis, is the following:
\begin{equation}\label{PSEUunr}
\rho(t)=\frac{N_{0}^{P}(t)}{N^{P}}\ket{\psi_{0}^{P}(t)}\bra{\psi_{0}^{P}(t)}+\frac{N_{1}^{P}(t)}{N^{P}}\ket{\psi_{1}^{P}}\bra{\psi_{1}^{P}},
\end{equation}
where
\begin{eqnarray}
&\ket{\psi_{0}^{P}(t)}=C_{g0}^{P}(t)\ket{g,0}+C_{g1}^{P}(t)\ket{g,1}+C_{e0}^{P}(t)\ket{e,0},\\
&\ket{\psi_{1}^{P}}=\ket{g,0},
\end{eqnarray}
$N^{P}$ is the total number of ensemble members, $N_{1}^{P}(t)$ is
the number of ensemble members who have decayed into the
atom-pseudomode ground state via a pseudomode jump, and
$N_{0}^{P}(t)$ is the number of members who have not jumped. If we
want to look at the time evolution of the ensemble members in the
atomic Hilbert space only, we have to trace out the pseudomode
auxiliary degree of freedom. This leads to the following reduced
atomic density matrix:
\begin{equation}\begin{split}\label{TraPSEUunr}
\rho_{A}(t)=&\frac{N_{0}^{P}(t)}{N^{P}}\Bigl((|C_{g0}^{P}(t)|^2+|C_{g1}^{P}(t)|^2)\ket{g}\bra{g}\\
+&|C_{e0}^{P}(t)|^2\ket{e}\bra{e}+C_{g0}^{P *}(t)C_{e0}^{P}(t)\ket{e}\bra{g}\\
+&C_{g0}^{P}(t)C_{e0}^{P
*}(t)\ket{g}\bra{e}\Bigl)+\frac{N_{1}^{P}(t)}{N^{P}}\ket{g}\bra{g}
\end{split}\end{equation}
in which the ensemble members are clearly in a mixed state. A
comparison between Eqs. \eqref{NMQJunr} and \eqref{TraPSEUunr}
illustrates the connection between the two unravelings. In
particular it is illustrative to consider the ground state
population,
\begin{equation}\begin{split}
\bra{g}\rho_{A}(t)\ket{g}&=\frac{N_{1}(t)}{N}+\frac{N_{0}(t)}{N}|C_{g}(t)|^2\\
&=\frac{N_{1}^{P}(t)}{N^{P}}+\frac{N_{0}^{P}(t)}{N^{P}}(|C_{g0}^{P}(t)|^2+|C_{g1}^{P}(t)|^2),
\end{split}\end{equation}
further showing the unravelings connection.

\subsection{The pseudomode as effective memory}
The interpretation of Eq.~\eqref{gammapseu1} is particularly
interesting in the light of the NMQJ method, where negative decay
rates lead to reversed quantum jumps. Consider, for example, an atom
initially prepared in a generic superposition of ground and excited
state performing a quantum jump to its ground state at a certain
time $t'$. If the decay rate $\g(t)$ becomes negative at $t > t'$,
the superpositions destroyed by the earlier normal jump can be
restored by a reversed jump. In fact a reversed jump takes the atom
to the state into which it would have evolved if the previous
quantum jump had not occurred. Thus in the NMQJ framework one can
characterize the period of negativity of the decay rate as the
period of time in which memory effects and restoration of quantum
superpositions occur through reversed quantum jumps. Stated another
way, the reverse jumps describe the process through which the system
recovers part of the information that leaked into the environment.
This is confirmed by the dynamics of the von Neumann entropy of the
atom which shows an oscillatory behavior following the oscillations
of the atomic decay rate $\g (t)$. In particular the atomic von
Neumann entropy decreases when the decay rate is negative indicating
a temporary reduction of the mixedness of the atomic state and a
recovery of coherence.

\begin{figure}[!]
\begin{center}
\includegraphics[width=8.6cm]{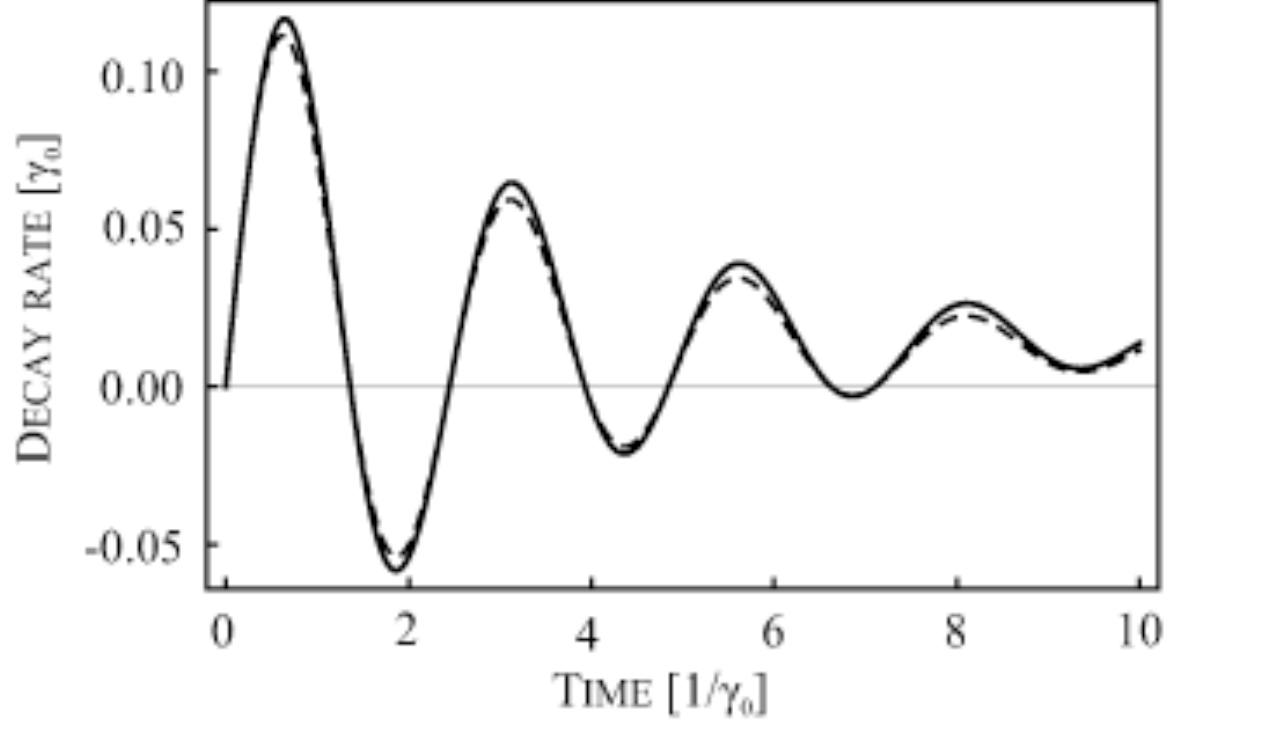}
\end{center}
\caption{The solid line is the decay rate for a two\,-\,level atom
in a Lorentzian structured reservoir. The dashed line is the
compensated rate of change of the pseudomode population
$d|b_{1}(t)|^2/dt+\G|b_{1}(t)|^2$. The units for the rates on the
vertical and horizontal axis are $\g_{0}$ and $1/\g_{0}$ with
$\g_{0}=4\Om_{0}^{2}/\G$ Markovian decay rate of the atom. We have
taken the values $\G=0.6\ \g_{0}$, $\Om_{0}=\sqrt{0.15}\ \g_{0}$ and
$\om_{c}-\om_{0}=4\G$.}\label{fig:grafico}
\end{figure}

Equation \eqref{gammapseu1} states that if the decay rate is
negative then the compensated rate of change of the pseudomode
population is negative as well, as clearly shown by
Fig.~\ref{fig:grafico}. So whenever the atom increases its excited
state population the pseudomode must deplete. This equation,
therefore, establishes a link between the restoration of coherence,
typical of a reversed jump, and the pseudomode depletion.
This observation suggests an interpretation of the pseudomode as
that part of the reservoir from which the atom receives back
information and probability due to memory effects. In other words
the pseudomode can be seen as an effective description of the
reservoir memory. This is further shown by the dynamics of the
mutual information \cite{qinfproc, Cerf} between atom and pseudomode
which perfectly follows the oscillation of the pseudomode
population. However the pseudomode is not a perfect storage place
since the efficiency of the information restoration from the
pseudomode to the atom depends on the pseudomode loss rate $\Gamma$.

\section{Memory structure and pseudomode architecture}
We now generalize our study to a more complicated spectral
distribution, namely the inverted Lorentzian mo-$\\$del describing
in a simplified way photonic band gaps \cite{Garr97}
\begin{equation}\label{twolor}
D(\om)=\frac{W_{1}\G_{1}}{(\om-\om_{c})^2+(\G_{1}/2)^2}-\frac{W_{2}\G_{2}}{(\om-\om_{c})^2+(\G_{2}/2)^2}.
\end{equation}
The negative Lorentzian introduces a dip into the density of states
leading to the inhibition of atomic spontaneous emission in the
region of the dip. In particular, for $W_{1}/\G_{1}=W_{2}/\G_{2}$,
the spectral distribution in Eq.~\eqref{twolor} presents a perfect
gap, $D(\om_{c})=0$. The exact pseudomode master equation is given
by \cite{Garrmath}
\begin{equation}\begin{split}\label{masteqpseu2}
\frac{d\rho}{dt}=&-i[H_{0},\rho]-\frac{\G'_{1}}{2}[a_{1}^{\dagger}a_{1}\rho-2a_{1}\rho
a_{1}^{\dagger}+\rho
a_{1}^{\dagger}a_{1}]\\&-\frac{\G'_{2}}{2}[a_{2}^{\dagger}a_{2}\rho-2a_{2}\rho
a_{2}^{\dagger}+\rho a_{2}^{\dagger}a_{2}],
\end{split}\end{equation}
where
\begin{equation}\begin{split}
H_{0}=&\om_{0}\s_{+}\s_-+\om_{c}a_{1}^{\dagger}a_{1}+\om_{c}a_{2}^{\dagger}a_{2}+\Om_{0}[a_{2}^{\dagger}\s_{-}+a_{2}\s_{+}]\\
&+V(a_{1}^{\dagger}a_{2}+a_{1}a_{2}^{\dagger}),
\end{split}\end{equation}
where $a_{1}$ and $a_{2}$ are the annihilation operators of the two
pseudomodes decaying with decay rates
$\G'_{1}=W_{1}\G_{2}-W_{2}\G_{1}$ and
$\G'_{2}=W_{1}\G_{1}-W_{2}\G_{2}$, respectively. The two pseudomodes
are coupled and $V=\sqrt{W_{1}W_{2}}(\G_{1}-\G_{2})/2$ is the
strength of the coupling. Figure~\ref{fig:schema}~(b) shows the
atom\,-\,pseudomodes architecture in this case. The atom interacts
coherently with the second pseudomode, which is in turn coupled to
the first one. Both pseudomodes are leaking into independent
Markovian reservoirs. In the case of a perfect gap only the second
pseudomode leaks. The set of ordinary differential equations
associated to the master equation \eqref{masteqpseu2} is
\begin{equation}\begin{split}\label{diffsyst4}
i\frac{d}{dt}c_{1}&=\om_{0}c_{1}+\Om_{0}a_{2},\\
i\frac{d}{dt}a_{1}&=z'_{1}a_{1}+V a_{2},\\
i\frac{d}{dt}a_{2}&=z'_{2}a_{2}+V a_{1}+\Om_{0}c_{1},
\end{split}\end{equation}
where $c_{1}$, $a_{1}$ and $a_{2}$ are the complex amplitudes for
the states with one excitation in the atom, one excitation in the
first pseudomode, and one excitation in the second pseudomode,
respectively. The position of the true poles is $z'_{1}=\om_{c}-i
\G'_{1}/2$ and $z'_{2}=\om_{c}-i \G'_{2}/2$.

Similarly to the calculations for the Lorentzian case one can show
that, after tracing out the two pseudomodes in
Eq.~\eqref{masteqpseu2}, and with the help of
Eqs.~\eqref{diffsyst4}, one obtains the non\,-\,Markovian master
equation \eqref{tracedME} for the atom, where $A(t)$ and $B(t)$ are
given by the same expressions in Eq.~\eqref{Asolution} and
\eqref{Bsolution} provided that we replace $b_{1}(t)$ with
$a_{2}(t)$.

The non\,-\,Markovian dynamics of the atom is linked to the coherent
variation of both pseudomodes by the following equation
\begin{equation}\label{gammapseu2}
\frac{d|a_{1}(t)|^2}{dt}+\G'_{1}|a_{1}(t)|^2+\frac{d|a_{2}(t)|^2}{dt}+\G'_{2}|a_{2}(t)|^2=\g(t)|c_{1}(t)|^2.
\end{equation}
This equation  generalizes Eq.~\eqref{gammapseu1} to the more
complex reservoir structure considered here.
Moreover, using Eqs.~\eqref{diffsyst4} we obtain a relation
connecting the dynamics of the first and second pseudomodes,
\begin{equation}\label{firstsecondpse}
\frac{d|a_{1}(t)|^2}{dt}+\G'_{1}|a_{1}(t)|^2=2
V\frac{\textrm{Im}\{a_{2}(t)a_{1}^{\ast}(t)\}}{|a_{2}(t)|^2}|a_{2}(t)|^2.
\end{equation}
Having in mind that
$\g(t)=2\Om_{0}\textrm{Im}\{c_{1}(t)b_{1}^{\ast}(t)\}/|c_1(t)|^2$
one sees that Eq.~\eqref{firstsecondpse} has the same structure as
Eq.~\eqref{gammapseu1}. In fact the compensated rate of change of
the first pseudomode population equals a time dependent coefficient
times the second pseudomode population. Therefore the first
pseudomode acts as a memory storage for the second one. Such a
storage of memory of the second pseudomode into the first one is
perfect in the case of perfect gap where $\G'_{1}=0$.

\section{Conclusion}
Both the pseudomodes and the reversed quantum jumps describe the
non\,-\,Markovian backaction of a reservoir on a quantum system. Our
results show that the picture of open quantum systems as comprised
of a system and environment is a commonly held view that does not
reflect the dynamics when the coupling is 'strong'; the environment
divides into memory and non-memory parts. Establishing a connection
in the case of a general structured reservoir, i.e., for a generic
form of the spectrum is an extremely complicated issue because both
the number of pseudomodes and their equations of motion depend on
the details of the reservoir structure. In this sense our results
constitutes a first step in the direction of an understanding of
memory in non-Markovian systems.

It is likely that a measurement on the reservoir will in general
affect any memory that it contains, and so far it has not been
possible to attach any measurement scheme to non\,-\,Markovian
reservoirs, as Ref. \cite{Gambetta08} demonstrates. However, since
we have seen that it is possible to make a separation of the
reservoir into a memory part and a Markovian reservoir part, there
remains the possibility that we could make a measurement scheme that
would address only the Markovian part of the reservoir. In the
cavity-atom situation, this might correspond to a measurement on
'outside'-modes, which have a significant amplitude outside the
cavity. In such a case it is clear that the reservoir-memory for the
atom actually resides in inaccessible 'inside' mode, or pseudomode,
belonging to the cavity.

\acknowledgements

This work has been supported by the Academy of Finland (Projects
No.~108699, No.~115682, No.~115982, and No.~8125004), the Magnus
Ehrnrooth Foundation, CIMO and Leverhulme Trust. S.M. also thanks
the Turku Collegium of Science and Medicine for financial support.

\end{document}